\documentclass[a4paper, twocolumn]{article}
\usepackage[pdftex]{graphicx,xcolor}
\usepackage{authblk}

\usepackage{url}

\title{Taxonomy of Risks on Automated Fact-Checking Systems Considering its Propagation}
\author[1]{Jun Yajima \thanks{jyajima@fujitsu.com}}
\author[1]{Tatsuya Oka}
\author[2]{Takao Okubo}

\affil[1]{Fujitsu Limited}
\affil[2]{Institute of Information Security}
\date{}

\begin{document}
\maketitle

\begin{abstract}
In recent years, the posting of fake news including disinformation and misinformation on social networking services (SNS) has become a social problem. To combat this fake news, fact-checking that is the process of assessing the veracity of posts on SNS has become increasingly important. While fact-checking is currently performed by fact-checking organizations, it is difficult to fact-check all posts on SNS. Therefore, the use of automated fact-checking systems is effective. Recent automated fact-checking systems utilize artificial intelligence and large language models, so there are risks of incorrect judgments and posting incorrect results on social media which can lead to the spread of misinformation or to engage in defamation.
In this paper, as a first step toward enabling the safe use of automated fact-checking systems, we categorize the specific risks on automated fact-checking systems. In this categorizing, we consider a three-stage risk propagation: risk factors, hazardous situations, and harm. Our analysis revealed that 32 specific risks exist in automated fact-checking systems. In this paper, we utilize the categorized risks as analytical cues (guide words) to present the risk assessment of the automated fact-checking system DEFAME. This assessment result indicates that risks that cannot be derived using STRIDE, a conventional IT security risk assessment method can be derived using our guide words. 
\end{abstract}

\section{Introduction}
\label{S-Introduction}
In recent years, posting and spreading fake news on social network services (SNS) is a major social problem. 
For example, at the 2016 U.S. presidential election, false rumors about a pizza restaurant spread and led to a shooting incident \cite{Pizzagate}. 
In Japan, whenever a massive earthquake strikes, posts claiming that the earthquake was artificially triggered, posts that stoke anxiety by sharing footage of past tsunamis even though no tsunami has actually occurred, and posts about earthquake prediction lacking scientific basis often go viral\cite{earthquake-nhk}.
These posts are made for a variety of reasons, including cases where the goal is to generate ad revenue through page views, potential disinformation campaigns originating overseas, and instances where those sharing the content genuinely believe the misinformation and spread it.
To avoid being misled by fake news, fact-checking is essential.
Fact-checking is carried out by fact-checking organizations such as PolitiFact\cite{politifact} in the United States, Full Fact\cite{fullfact} in the United Kingdom, and the Japan Fact-Checking Center\cite{JFC} in Japan.
While fact-checking organizations accurately verify the truthfulness of major posts, not all posts on SNS are fact-checked, and social media users cannot immediately judge the truthfulness of every post they see.
Therefore, automated fact-checking systems which can perform fact-checking automatically are expected to serve as a complementary tool to the fact-checking conducted by fact-checking organizations.
Research into automated fact-checking technology is ongoing, and technological development is progressing. 
For example, the development project of automated fact-checking systems in Japan is being conducted\cite{FJ-Press}.
Applications of automated fact-checking systems include use by analysts in fact-checking organizations, use by corporate and other organizational users to identify dis/misinformation targeting their organizations, and use by social media users to fact-check posts they encounter.

Since automated fact-checking systems are typically implemented using artificial intelligence (AI) and large language models (LLMs), there are concerns about various risks associated with these technologies. 
For example, users seeking to generate ad revenue might input sensational disinformation into an automated fact-checking system and repeatedly tweak the text little by little until the system classifies the claim as ``true.'' 
If they can obtain their disinformation that is deemed ``true'', they use it as an endorsement to try to generate a large number of impressions. 
For example, take the false claim that ``Yesterday's earthquake was artificially triggered.'' 
By cleverly manipulating (or modifying) the input text and feeding it into the system repeatedly, the attacker continues until a ``true'' result is obtained.
Due to the variability and hallucinations in AI and LLM results, if even a single instance among the vast amount of input data yields a ``true'' result, 
there is a risk that disinformation--such as ``The claim that the earthquake was artificial is correct. This automated fact-checking systems have verified it as true''--
will be posted and spread using the fact-checking results as an endorsement. 
From the perspective of developers/providers of automated fact-checking systems, it is necessary to identify the risks associated with such systems in advance and consider appropriate measures before releasing them.

The following are existing technologies and initiatives regarding the risks of automated fact-checking systems.
Pennycook et al. analyzed fake news from a psychological perspective, examining how fake news comes to be believed and which types of topics are more likely to be believed\cite{Pennycook}. This paper points out that automated fact-checking faces misjudgment. 
However, harm of such risks in normal use or abuse of automated fact-checking systems are not described.
STRIDE\cite{STRIDE} is frequently used to identify IT security risks, 
but we argue that identifying risks associated with recent machine learning, generative AI, and risks specific to automated fact-checking are difficult. 
Because STRIDE was established and widely used before these technologies became widespread. 
Nakov et al. introduces initiatives in automated fact-checking and describes challenges related to its realization \cite{Nakov}. 
Guo et al. provides a survey of automated fact-checking, offering a technical overview of the field and introducing relevant datasets \cite{Guo}. 
However, these papers do not address the potential risks or misuse associated with automated fact-checking.
The Japan AI safety institute (AISI) published a report\cite{AISI-KA} summarizes attacks on AI systems and their consequences. 
While this information is useful, the descriptions are limited to direct damage such as output information leakage and model malfunction; 
they do not address the indirect harm suffered by users who receive the results of a malfunctioning model, nor do they cover further harm resulting from the misuse of those results. 
This is insufficient from the perspective of comprehensively covering the risks of automated fact-checking.
The Poynter website\cite{poynter} lists problems in fact-checking, but these problems are organized from the perspective of human fact-checkers and do not address the risks of automated fact-checking.
Tanaka et al.'s categorization of generative AI risks\cite{Tanaka} organizes generative AI risks into 20 categories and can be used as a reference for our own risk categorization.
The AISI Japan's AI Safety Evaluation Perspectives Guidelines \cite{AISI-SG} list categories of AI safety and are a useful reference. 
However, it does not provide detailed information on the risks involved in automated verifying the truthfulness of mis/disinformation.
So, we took this guide only as a reference information.

This paper identifies the risks specific to automated fact-checking systems with the aim of helping developers/providers systematically understand their risks.
This paper excludes general IT security and AI security from its scope and focuses on discussing risks specific to automated fact-checking systems.
In this analysis, we consider a three stage chronological sequence--risk factors, hazardous situations, and harm--and organize the findings into tables of risks.
After that, to confirm the appropriateness of the tables, we use each of these identified risks as analytical cues (guide words) to extract risks associated with the automated fact-checking system DEFAME and discuss the appropriateness of result.
Our analysis clarified that we could categorize the risks on automated fact-checking systems into 32 types and clarified these risks could be used as guide words to identify risks in automated fact-checking systems (e.g., DEFAME.) 

This paper is organized as follows.
In Section \ref{S-FakeNews}, we introduce fake news and explain the importance of automated fact-checking systems and the risks associated with them.
In Section \ref{S-RelatedWork}, we review prior research related to our analysis.
In Section \ref{S-Risk}, we categorize the risks of automated fact-checking systems and describe a fault tree of the logical relationships between each risk.
In Section \ref{S-Evaluation}, we present the results of an analysis of DEFAME’s risks based on the risk classification results.
In Section \ref{S-Discussion}, we discuss the risk classification presented in this paper.
In Section \ref{S-Conc}, we provide a summary and future works.

\subsection{Contributions of this paper}
The contributions of this paper are as follows. 
\begin{enumerate}
\item Identified 32 risks specific to automated fact-checking systems as a basic study of risks on them
\item Organized the identified risks considering chronological sequence (risk factors, hazardous situations, harm) and represented them into a fault tree (enabling the tracing of logical relationships)
\item A case study demonstrated that risks can be systematically derived on a Data Flow Diagram (DFD) through risk assessment using the 32 identified risks as guide words
\item A comparison with STRIDE demonstrated that social impacts (defamation, spreading of disinformation, etc.) and other risks can be assessed that cannot be assessed by STRIDE. 
\end{enumerate}

\section{Fake news and automated fact-checking system}
\label{S-FakeNews}
\subsection{Fake news}
\label{SS-FN-FakeNews}
The main purposes of generating and spreading fake news are as follows: 
\begin{itemize}
\item Financial motives

Since ad revenue can sometimes be earned based on the number of views or clicks (impressions) of content on SNS, fake content may be created for financial gain.

\item Political motives (attacks on the social system through incitement of the masses)

Fake content may be created with the aim of damaging the reputation of an opponent to benefit a supported political party or candidate during political events such as elections.

\item Attacking indivisuals or organizations

Fake content may be created to damage the reputation of disliked celebrities or companies and undermine their credibility.

\item Hedonistic motives

Fake content may be created simply to derive satisfaction from spreading disinformation.

\end{itemize}

Some well-known examples of fake news include the following:

\begin{itemize}
\item The Pizzagate Incident

Suspicions of a pizza shop was involved in criminal activity led to a shooting incident\cite{Pizzagate}.

\item The lion hoax during the Kumamoto Earthquake in Japan

A hoax claiming that a lion had escaped from a zoo was posted, causing severe anxiety among disaster victims. 
The poster was arrested for obstruction of business by fraudulent means\cite{kumamoto}.

\end{itemize}
These examples show that the spread of fake news has serious negative effects on society and making countermeasures are necessary.

\subsection{Fact-checking}
\label{SS-FN-FackCheck}
Fact-checking is one of the measures taken to combat fake news.
Fact-checking is used for verifying the truthfulness of social media posts and other content by consulting the reliability of evidence information obtained from official and/or other sources.

Fact-checking is conducted by organizations such as PolitiFact, FullFact, and Japan Fact Check Center (JFC), and findings of several impactive posts on SNS are published. 
While fact-checking by these organizations is reliable, the scope of their checks is limited to posts that are presumed to have a significant impact or that have been widely shared.
Therefore, from the perspective of social media users, it is difficult to verify the truthfulness regarding of all posts they encounter.
One solution to this problem is to use automated fact-checking systems that perform fact-checking via IT systems.
Automated fact-checking systems automatically verify the truthfulness of inputted posts from various angles and output fact-checking results.
Well-known examples of automated fact-checking systems are DEFAME\cite{DEFAME} and FacTool\cite{FacTool}.

\subsection{Potential risks on automated fact-checking systems}
\label{SS-FN-Risks}
While automated fact-checking systems are effective against fake news, we think they have risks specific to them.
For example, since automated fact-checking systems are typically implemented using artificial intelligence (AI) and large language models (LLMs), there is a risk of misjudgment caused by various technical problems (e.g., lack of accuracy, training data quality, or hallucination.) 
If an automated fact-checking system mistakenly classifies mis/disinformation as ``true'' due to technical problems, it could mislead users; in addition, if users post the results on social media, this could lead to the spread of mis/disinformation. 
Furthermore, there is a risk that attackers seeking to spread their disinformation could repeatedly input fake news made by them into the automated fact-checking system until it yields a ``true'' result and then combine them into a single post stating that ``This news is true. Truthfulness is guaranteed by fact-checking''. 
In this scenario, the automated fact-checking system would effectively be endorsing the attacker’s fake news and facing a risk that the system could inadvertently facilitate the spread of disinformation. 
The automated fact-checking systems are subject to various specific risks.
In this paper, we categorize the risks associated with automated fact-checking systems by referencing existing risk assessment techniques and risk classifications. 
We then use these categorized risks as guiding principles to identify the risks of the automated fact-checking system DEFAME and verify the validity of our risk classification.

\section{Related works}
\label{S-RelatedWork}
\subsection{STRIDE}
\label{SS-RW-STRIDE}
STRIDE is a technique widely used in security risk assessments in the IT field\cite{STRIDE}. 
In STRIDE, creating a data flow diagram (DFD) and assessing risks on the DFD.
Each letter -- S, T, R, I, D, and E -- represents a specific type of risk. 
These types of risks are used as guide principles (guide words) for assessing where and what kinds of risks may be hidden at the data boundaries and entities depicted in the DFD. 
The meanings of each letter in STRIDE are as follows:
\begin{itemize}
\item S: Spoofing
\item T: Tampering
\item R: Repudiation
\item I: Information disclosure
\item D: Denial of service
\item E: Elevation of privilege
\end{itemize}
STRIDE is a security risk assessment technique for IT systems that has been in use for many years.
Automated fact-checking systems mostly be implemented using machine learning (ML) and LLMs, ML and LLMs have only recently been put into practical use, making it difficult for traditional technique STRIDE to identify risks associated with them.
Furthermore, STRIDE cannot identify risks specific to automated fact-checking systems easily. 
This is discussed in Section \ref{SS-D-STRIDE}.

\subsection{Risk classification for machine learning}
\label{SS-RW-ML}
In 2021, the authors classified the types of risks in ML as part of an initiative independent of this study (not published).
In this classification, we added guide words that were missing from STRIDE for the risk assessment of ML systems and removed guide words that were deemed unnecessary.
The results of this reorganization of the guide words are as follows.

\begin{itemize}
\item Abuse
\item Adversarial Tampering
\item Denial of Service
\item Plagiarism
\item Unauthorized Use
\item Poisoning
\item Information Disclosure
\item Inference
\item Spoofing
\end{itemize}
These can be used in place of STRIDE as guide words for risk assessment of ML systems. 
However, since this reorganization was conducted in 2021, it cannot capture the risks associated with LLMs and generative AI --such as ChatGPT, whose use has expanded since then.

\subsection{Risk Taxonomy of generative AI}
\label{SS-RW-Tanaka}
Tanaka et al. categorized the risks associated with generative AI into 20 types\cite{Tanaka}. 
This categorization organizes the risks of generative AI with reference to ISO/IEC Guide 51\cite{ISO51}. 
In Tanaka et al.'s paper, they organize risks considering the chronological sequence from their origin to their impact. 
Although these risks are specific to generative AI and cannot be directly applied to risks of automated fact-checking systems, 
we decided to use them as a reference because automated fact-checking systems mostly utilize generative AI. 
When structuring the risks of automated fact-checking systems, we also consider the chronological sequence, as in ISO/IEC Guide 51.
The specific risk items identified by Tanaka et al. will be introduced in Section \ref{SS-R-Risk} as explanation of correspondence between Tanaka's risk items and our work of risks of automated fact-checking systems.

\subsection{Evaluation Perspecties on AI Safety}
\label{SS-RW-AISI}
The Japan AISI has published a guide to evaluation perspectives on AI safety based on the concept of AI safety, which addresses various risks associated with AI\cite{AISI-SG}.
This guide lists 10 items related to AI safety. 
This guide includes an item on preventing mis/disinformation. 
However, it does not provide detailed information on the risks involved in automated verifying the truthfulness of mis/disinformation.
So, we use this guide as only reference information. 
Their evaluation perspectives on AI safety are as follows. 
\begin{itemize}
\item Control of Toxic Output
\item Prevention of Misinformation, Disinformation and Manipulation
\item Fairness and Inclusion
\item Addressing to High-risk Use and Unintended Use
\item Privacy Protection
\item Ensuring Security
\item Explainability
\item Robustness
\item Data Quality
\item Verifiability
\end{itemize}

\section{Risks on Automated Fact-Checking Systems}
\label{S-Risk}
\subsection{Identifying Risks on Automated Fact-Checking Systems}
\label{SS-R-Risk}
As described in Section \ref{SS-RW-STRIDE}, using STRIDE alone is difficult to identify automated fact-checking systems' specific risks. 
The reasons are discussed in Section \ref{SS-D-STRIDE}. 
We organize the risks specific to automated fact-checking systems by referencing the risk issues described in Tanaka et al.'s paper \cite{Tanaka}, as introduced in Section \ref{SS-RW-Tanaka}.
Specifically, we considered the correspondence between the 20 risk issues identified in their paper and how they appear in the automated fact-checking systems.
The following outlines the relationship and consideration to organizing risks in automated fact-checking systems based on the risk issue in their paper.
We also considered the ML risk classification in Section \ref{SS-RW-ML} and the Japan AISI evaluation guide in Section \ref{SS-RW-AISI} as reference information. 

\begin{enumerate}
\item Hallucination

We classified ‘‘hallucinations’’ as "misjudgments’’ in a broad sense. A misjudgment of AI occurs when some kind of misjudgment factors (e.g., hallucinations, training data bias) arise in the normal operation of fact-checking or it intentionally occurred by adversarial attacks. 
This can lead to the spread of misinformation, defamation, and the generation of impressions on social media for the purpose of generating ad revenue.

\item Potential for risky emergent behaviors

This issue relates to misjudgment in normal use or intentional misjudging by adversarial attacks.

\item Harmful content

This relates to the system producing harmful output. 
For example, this involves producing various harmful outputs, such as violating laws or encouraging criminal activity. 
Alternatively,  the automated fact-checking system that conducts malicious judgments or decisions that favor specific individuals or organizations.
This could result in unfair judgments, which in turn could lead to the spread of mis/disinformation, defamation, and the generation of impression-based revenue.

\item Harm of representation, allocation, and quality of service

We think that this issue relates to the situation that automated fact-checking systems may cause misjudgment. We also think this issue relates to item of harmful content. 

\item Disinformation and influence operations

This is similar to cases of misjudgment and could lead to the spread of mis/disinformation, defamation, and the pursuit of page views for advertising revenue.

\item Overreliance

Users believe the system's results are always correct, leading them to make incorrect judgments or spread mis/disinformation when misjudgment occurs.

\item Privacy

In appropriate provision of privacy information used in fact-checking, such as supporting evidence and assessment results, may result in privacy violations. 
This also involves information disclosure by adversarial attacks. 

\item Copyright infringement

In case of the evidence or assessment results contain copyrighted material, which may result in copyright infringement. 

\item Exploitation of workers during model creation

Workers who remove harmful data may suffer physical and mental harm as a result. 

\item Cybersecurity

This relates to an attack designed to intentionally cause misjudgments or privacy disclosure. 
It has the potential to lead to data leaks, the spread of mis/disinformation, defamation, and the generation of impressions for the purpose of generating ad revenue.
By repeatedly accessing the system, attackers could steal training data or hijack models, or they could inject their own data into the training data to cause misjudgment.

\item Economic impacts

Use of automated fact-checking systems will lead to staff cuts among fact-checkers and cause financial hardship for those affected

\item Acceleration

This relates to acceleration of competition with rival systems. 
As technological competition intensifies, 
there is a risk that companies may report quality as being better than it actually is, even when there is no factual basis for such claims.
This raises the risk of incorrect judgments due to inadequate quality of system. 
This also relates to overreliance. 

\item Environmental and financial cost

There are concerns about power consumption and its cost when using AI. 

\item Spreading misinformation

If the system makes a misjudgment in normal use or case of adversarial attacks, misinformation is spread.

\item Increasing sophistication and ease of crime

By having the system determines the validity of the crime methods, 
making criminal methods easier to commit crimes and lead to more sophisticated methods.

\item Proliferation of conventional and unconventional weapons

Entering instructions to investigate the method of making weapons into fact-checking tools could contribute to the proliferation of weapons. This could lead to human damage or legal violations.

\item Illegal surveillance and censorship

We don't think this issue relates to automated fact-checking systems' risks

\item Lack of transparency of training data

A lack of explanation regarding the evidence or results may lead to misunderstandings among users.

\item Interactions with other systems

This issue relates that defamation will be posted on SNS when results of development system differ from other fact-checking systems

\item No rights (copyrights or patents) for AI

No copyright can be claimed over the judgment results. 

\end{enumerate}

\subsection{Taxonomy of Risks on Automated Fact-checking Systems}
\label{SS-R-Taxonomy}
We summarize the risks of automated fact-checking systems considered in Section \ref{SS-R-Risk} in Tables \ref{T-GuideWord1} and \ref{T-GuideWord2}. 
We organize risks chronologically into three stages-- ``Risk Factor,'' ``Hazardous situation,'' and ``Harm''-- by referring ISO/IEC guide 51. 
Table \ref{T-GuideWord1} lists risks that may occur during normal use of an automated fact-checking system. 
For example, a possible chronological risk scenario is as follows: 
if the AI used in the system makes a misjudgment due to errors caused by any factors such as inaccuracies or hallucinations in AI   (Mc) (risk factor), 
it results in a non-malicious misjudgment (Mj) (Hazardous situation) of final judgment of the system; 
users who trust the result then post the incorrect judgment on social media, thereby spreading misinformation (Fr) (harm).
On the other hand, Table \ref{T-GuideWord2} lists risks associated with scenarios where the automated fact-checking system is attacked. 
For example, if an attacker inputs malicious data (AMI) (risk factor) or poisons the evidence information on the web (AEP) (risk factor), 
the system may make a misjudgment (AMj) (Hazardous situation), 
and then the attacker posts this result on social media, 
it could lead to a chronological risk scenario where the attacker generates monetary income by gaining impressions on social media (AIm) (harm) and spreads disinformation (AFr) (harm).

\begin{table*}[t]
\caption{Risks of automated fact-checking systems in normal use}
\label{T-GuideWord1}
\centering
\begin{tabular}{|l|l|l|} \hline
Phase of risk & Risk name & Explanation \\ \hline
Risk factor & Misconstruction (Mc) & Misconstruction of system (design, implementation, or \\ 
& & errors caused by AI) (Insufficient training data, data \\ 
& & bias, hallucinations, etc.) \\ \hline
Risk factor & Input data (Ip) & Risks from input data (entry of fabricated data, or \\ 
& & prohibited data, etc.) \\ \hline
Risk factor & External model (Em) & Risks from external models or external data \\ \hline
Risk factor & Normal operation (No) & Risks in normal operations \\ \hline
Hazardous & Misjudgment (Mj) & Unintentional misjudgments or malfunctions \\
situation &  &  \\ \hline
Hazardousn & Unprecise explanation (Ue) & Discrepancies, gaps, or lack of transparency in the \\ 
situation & & explanation of evidence or results \\ \hline
Hazardous & Prohibited / inappropriate & Generate prohibited or appropriate outputs \\ 
situation & output (Po) &  \\ \hline
Hazardous & Unfairness output (Uf) & Generate biased results (favoring specific groups or  \\ 
situation &  & individuals) \\ \hline
Harm & Defamation (Dm) & Leads to defamation \\ \hline
Harm & False rumors (Fr) & Spreading Mis/disinformation \\ \hline
Harm & Maximize reach on social & Generating impressions for the purpose of ad revenue \\ 
 &  media (Mr) &  \\ \hline
Harm & Information disclosure (Id) & Information disclosure, privacy violations \\ \hline
Harm & Over confidence (Oc) & Believing results unquestioningly, over-confidence on \\
& & the system \\ \hline
Harm & Infringement of rights (Ir) & Infringement, Plagiarism \\ \hline
Harm & Legal violation (Lv) & Violations of or encouragement of laws and regulations \\ \hline
Harm & Encouraging crime (Ec) & Encouraging crime \\ \hline
Harm & Real-world damage (Rd) & Actual harm (physical, financial, material, or emotional \\
& & damage) \\ \hline
\end{tabular}
\end{table*}

\begin{table*}[tbp]
\caption{Risks of automated fact-checking systems by adversarial attacks}
\label{T-GuideWord2}
\centering
\begin{tabular}{|l|l|l|} \hline
Phase of risk & Risk name & Explanation \\ \hline
Risk factor & Adversarial Malicious  & Malicious input or entering prohibited input \\ 
 &  Input (AMI) &  \\ \hline
Risk factor & Adversarial Evidence  & Tampering with data used as evidence to make \\
 &  Poisoning (AEP) & it malicious \\ \hline
Risk factor & Adversarial Attacks (AA) & Launch adversarial attacks against AI systems to \\
& & cause misjudgment or data leaks \\ \hline
Hazardous  & Adversarial & Leads to misjudgment \\ 
situation & Misjudgment (AMj) & \\ \hline
Hazardous  & Adversarial Unprecise  & Leads to misunderstand of explanation \\ 
situation & explanation (AUe) & \\ \hline
Hazardous  & Adversarial Prohibited / & Leads to a prohibited or inappropriate output \\ 
situation & inappropriate output (APo) &  \\ \hline
Hazardous  & Adversarial Unfairness & Leads to a unfairness output \\ 
 situation & output (AUf) &  \\ \hline
Harm & Adversarial Defamation (ADm) & Leads to defamation \\ \hline
Harm & Adversarial False rumors (AFr) & Spreading Mis/disinformation \\ \hline
Harm & Adversarial Maximize  & Generating impressions for ad revenue \\ 
Harm & reach on social media (AMr) &  \\ \hline
Harm & Adversarial Information & Leads to information disclosure, or privacy  \\ 
Harm & disclosure (AId) & violations \\ \hline
Harm & Adversarial Infringement & Leads to infringement, or plagiarism \\ 
Harm & of rights (AIr) &  \\ \hline
Harm & Adversarial Legal & Leads to violations of laws and regulations or \\
&  violation (ALv) & encouraging such violations \\ \hline
Harm & Adversarial Encouraging & Leads to the encouragement of crime \\ 
Harm & crime (AEc) &  \\ \hline
Harm & Adversarial Real-world & Leads to actual harm (physical, financial, \\ 
Harm & damage (ARd) & material, or emotional damage)  \\ \hline
\end{tabular}
\end{table*}

\subsection{Fault tree}
\label{SS-R-FT}
\begin{figure}[htbp]
\centering
\includegraphics[width=\linewidth]{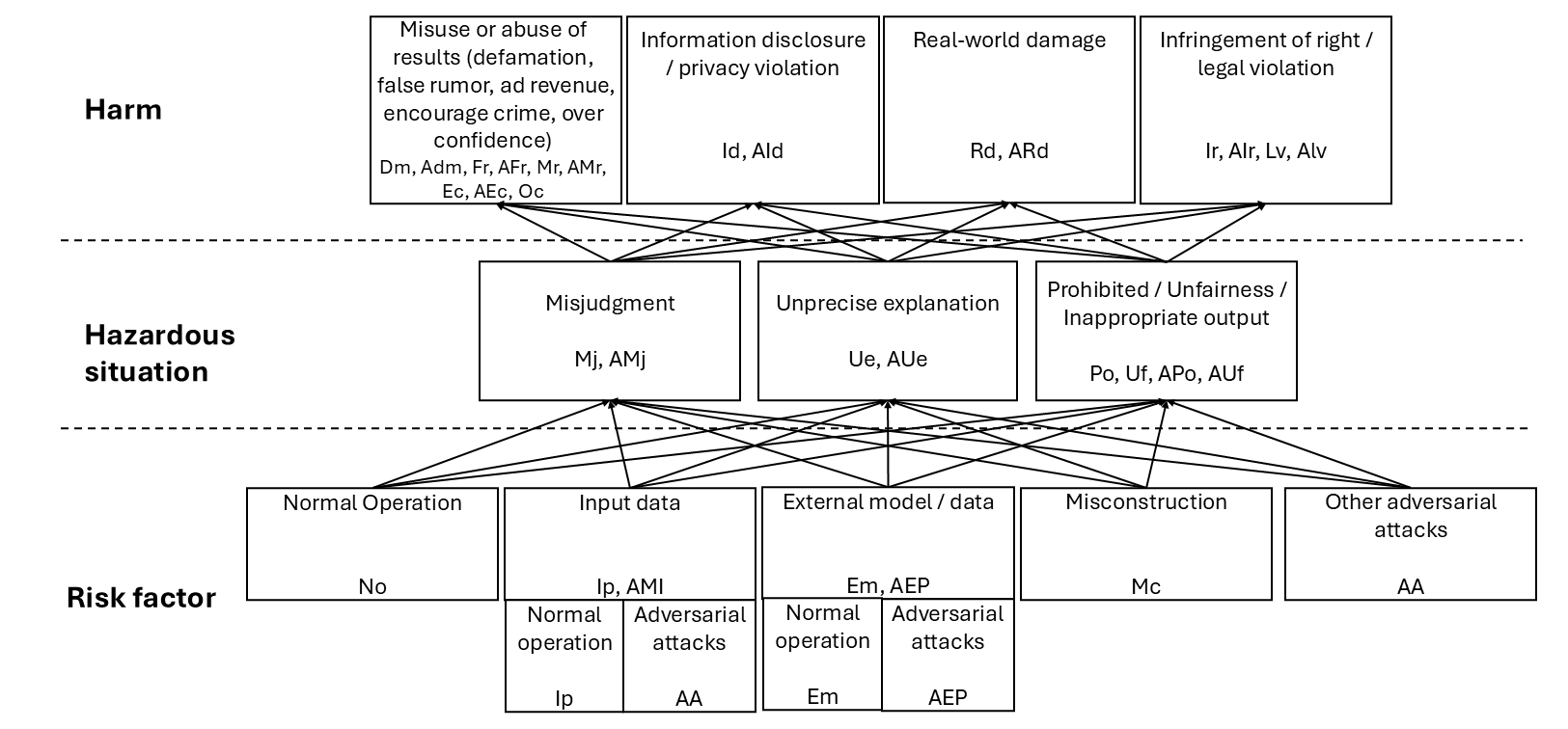}
\caption{Fault tree on automated fact-checking systems}
\label{F-FT}
\end{figure}

The tables of risks on automated fact-checking, organized in Section \ref{SS-R-Taxonomy} also show the chronological relationships between risks, including risk factors, hazardous situations, and harm. 
Figure \ref{F-FT} illustrates these chronological relationships as relationships in which cause and harm influence each other as a fault tree (FT)\cite{FTA}. 
The risks on upper nodes of Figure \ref{F-FT} arise when lower nodes are triggered. 
For example, the risk of defamation arises when the layer of hazardous situation, namely, misjudgment, unprecise explanation, or prohibited/unfairness/inappropriate output are triggered, and the hazardous situations are triggered when one of the risk factors are triggered. 
In other words, the risks on the upper layer cannot be triggered unless one of the risks on the lower layer is triggered. 
This figure helps to understand the logical structure of the risks that connect the risk factors to the harm via the hazardous situations.

\section{Evaluation on actual automated fact-checking system}
\label{S-Evaluation}
\subsection{Overview}
\label{SS-E-Overview}
In this section, to verify the validity of the tables of risks summarized in Section \ref{SS-R-Taxonomy}, 
we assess the risks associated with an actual automated fact-checking system. 
We use each of the identified risks as a guide word. 
Specifically, we identify the risks on the Dynamic Evidence-based FAct-checking with Multimodal Experts (DEFAME)\cite{DEFAME}, one of the well-known automated fact-checking systems.
Similar to the risk assessment using STRIDE, we draw a Data Flow Diagram (DFD) and assess whether the risks summarized in Section \ref{SS-R-Taxonomy} exist at each entity and data boundary on the DFD, thereby observing the validity of the risk taxonomy.

\subsection{DEFAME}
\label{SS-E-DEFAME}
DEFAME is an open-source automated fact-checking system\cite{DEFAME}. 
Based on the characteristics of the target claim, this system uses an LLM to autonomously select and utilize multiple entities such as web search and image geolocation, to gather evidence for determining the truthfulness of the claim. 
After that, LLM analyzes this evidence and presents the fact-checking result with the evidence for its conclusion.
DEFAME consists of six steps: Plan, Execute, Summarize, Develop, Judge, and Justify. 
When the data to be evaluated is input, the Plan step determines a strategy of which tools (Web Search, Image Search, Reverse Image Search, Geolocation, etc.) to use for fact-checking. 
Following the determined strategy, the Execute step collects evidence using the selected tools. 
Next, in the Summarize step, the results from each tool are summarized, and in the Develop step, the claim and evidence are integrated. 
In the Judge step, the system determines which category the claim belongs to, and depending on the situation, the process from first step may be repeated up to three times. 
In the Justify step, key findings and relevant evidence are summarized and presented in a format that is easy for humans to understand.

\subsection{Risk Assessment Case Study}
\label{SS-E-RiskAssessment}
\subsubsection{Assessment Procedure}
\label{SSS-E-RA-P}
\begin{enumerate}
\item Draw a data flow diagram (DFD)

Draw a DFD for the assessment target automatic fact-checking system.
Various types of evidence collection tools, judgment function, entities such as users, functions, tools, and boundaries of data exchange between entities should be depicted on DFD. 
The DFD we drew for DEFAME is shown in Figure \ref{F-DFD}.
DEFAME judgment process is executed in the DEFAME pipeline box. 
Since the evidence collection executed in the execute step uses external tools, they should be placed outside of the system boundary (on the right side of Figure \ref{F-DFD}). 
The LLMs used by the DEFAME pipeline and external tools are generally developed externally vendors. 
So, they should be placed outside of the boundary. 
Similarly, the search engine executed by the external tools should also be placed outside of the boundary.

\begin{figure}[htbp]
\centering
\includegraphics[width=\linewidth]{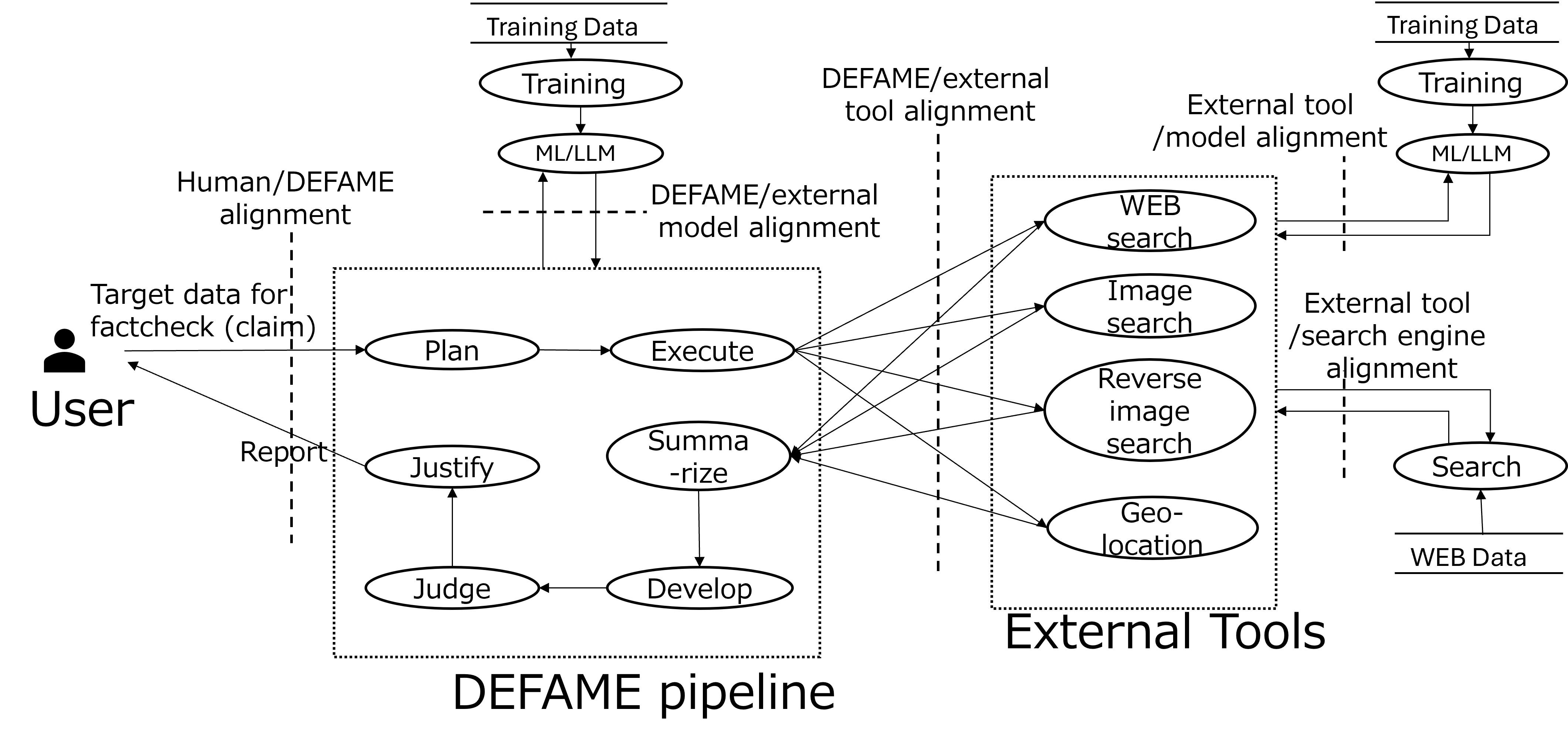}
\caption{Data Flow Diagram (DFD) on DEFAME}
\label{F-DFD}
\end{figure}

\item Identifying risks at entity and data boundaries on the DFD

For each entity and data boundary in the DFD created in the previous step, identify which of the risks listed in Table \ref{T-GuideWord1} and Table \ref{T-GuideWord2} occur. 
The result of the risk identification is shown in Figure \ref{F-DFD-risk}. 
In this figure, risks that occur in normal use (from Table \ref{T-GuideWord1}) are shown in italics, and risks resulting from adversarial attacks (from Table \ref{T-GuideWord2}) are underlined. 
Additionally, for comparison, the result of the STRIDE assessment is shown in bold.
The meanings of each abbreviation are shown in Table \ref{T-GuideWord1}, Table \ref{T-GuideWord2}, and Section \ref{SS-RW-STRIDE}. 
Among the risks identified in this paper, those related to ``harm'' include cases that third parties are affected when DEFAME users post fact-checking results on social media. 
For this reason, we decided to illustrate the posting on social media and the harmed third parties of the posts in the lower-left corner of Figure \ref{F-DFD}, as shown in Figure \ref{F-DFD-risk}.

\begin{figure*}[t]
\centering
\includegraphics[width=44em]{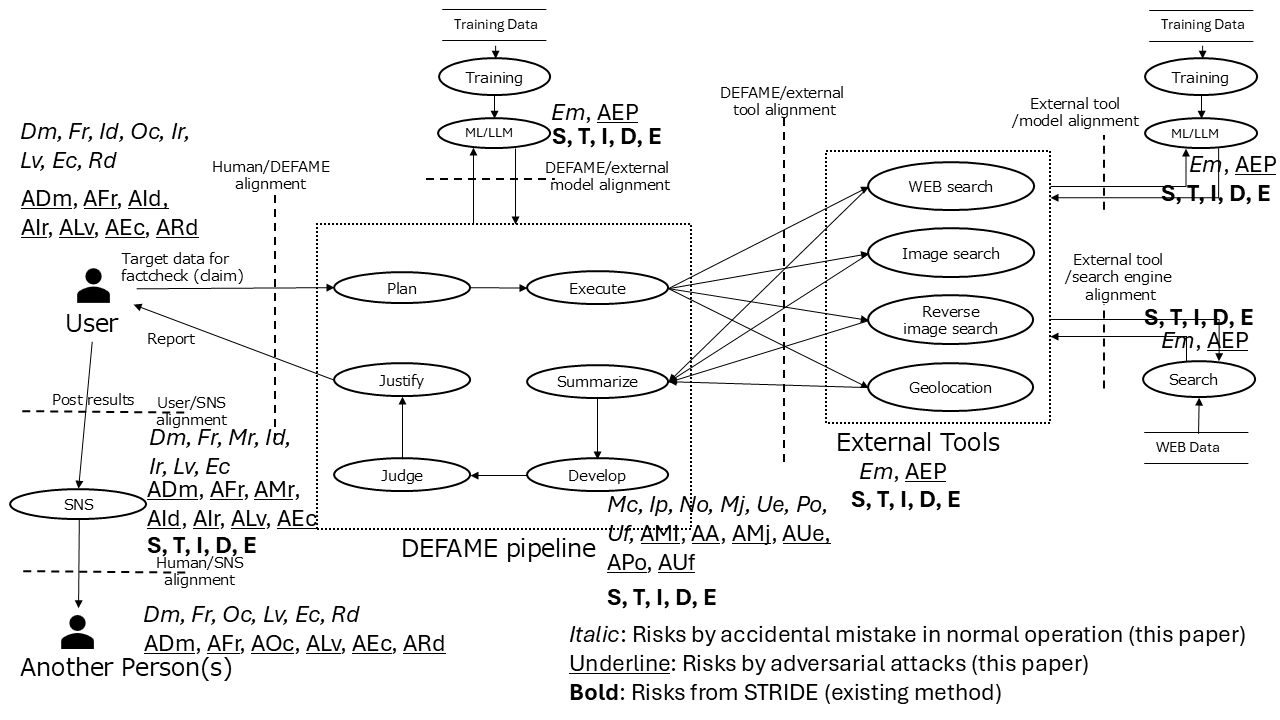}
\caption{Result of risk assessment on DEFAME}
\label{F-DFD-risk}
\end{figure*}

\end{enumerate}

\subsection{Observation}
\label{SS-E-Observation}
Looking at the result of the case study conducted in Section \ref{SS-E-RiskAssessment}, 
we found that among the risks categorized in this paper, 
the risk factors are located in the DEFAME pipeline, external tools, ML/LLM, and search engines. 
Furthermore, we found that hazardous situations exist in the DEFAME pipeline. 
This indicates that hazardous situation arises in the DEFAME pipeline after an incident occurs derived by risk factors that are in the system (including DEFAME pipeline, external tools, external ML models, external web search, or their boundaries.) 
Regarding actual harm, we found that they occur among DEFAME users, on social media if users post results on SNS, and among third parties by posting of the results. 
On the other hand, for STRIDE, we found that threats other than Repudiation occur at all entities and data boundaries except for users. 
This indicates that, since DEFAME is composed of general IT systems, general IT threats exist in the system. 
By this result, we can summarize that the risks cannot be identified by STRIDE such as the harm of the spread of defamation and mis/disinformation can be identified by using our taxonomy.

\section{Discussion}
\label{S-Discussion}
\subsection{Risk Assessment using STRIDE vs Ours}
\label{SS-D-STRIDE}
As shown in Section \ref{SS-E-Observation}, 
the identified risks using STRIDE and ours are differ. 
The main reason is that the target risks of each guide word are different. 
The differences between the STRIDE guide words and our guide words are shown in Table \ref{T-STRIDE-GuideWord}.
The target of STRIDE guide words is to identify general security risks on IT systems. 
In other words, identifying risks not associated with general security risks is difficult only using STRIDE guide words. 
On the other hand, our guide words' target is to identify automated fact-checking systems specific risks. 
By only using our guide words, though we can identify automated fact-checking system specific risks, it is difficult to identify general security threats of the systems. 
For example, a threat that tampering system output label and posting it on social media corresponds to the ``T: tampering'' in STRIDE. 
Since this threat can be identified using STRIDE, it is not included in our guide words.
By using both methods, it is possible to effectively assess risks of automated fact-checking systems for both aspects.

\begin{table*}[tbp]
\caption{Difference of our risk taxonomy and STRIDE}
\label{T-STRIDE-GuideWord}
\centering
\begin{tabular}{|c|c|c|} \hline
 & STRIDE & Risk taxonomy of this paper  \\ \hline
Target system & General IT systems including & Autometed fact-checking systems \\ 
 & automated fact-checking systems &  \\ \hline
Identified risks & General security threats & Harm on users, and the risk factors \\ 
 &  & and hazardous situations behind them \\ \hline
\end{tabular}
\end{table*}

\subsection{Comprehensiveness of risks}
\label{SS-D-Compre}
In this paper, we have identified risks specific to automated fact-checking systems in relation to the risks of generative AI described in Tanaka et al.'s paper \cite{Tanaka} and shown them in Table \ref{T-GuideWord1} and Table \ref{T-GuideWord2}. 
In this mapping process, we also identified risks with reference to the Japan AISI's guide \cite{AISI-SG} and the list of ML risks described in Section \ref{SS-RW-ML}. 
Therefore, if there are any lacks in the risks of Tanaka et al.'s paper, there will also be lacks in the risks listed in this paper. 
If there are risks specific to automated fact-checking systems that are unrelated to generative AI, then the list of risks provided would be incomplete. 
However, we believe that the generative AI risks identified in Tanaka et al.'s paper sufficiently cover the risks of ML systems that utilize generative AI. 
Because the risks pointed out in Tanaka et al.'s paper also include the risks of the ML used to implement generative AI, and we consider that they broadly cover the risks of ML systems implemented using ML and generative AI. 
Even if there is no guarantee that Tanaka et al.'s paper truly covers all risks associated with ML, including generative AI, the list of risks identified in this paper is still considered useful for developers and distributors of automated fact-checking systems when considering the risks specific of the systems they are developing or distributing. 

\subsection{Magnitude of Risk}
\label{SS-D-Severity}
This paper has identified various risks specific to automated fact-checking systems. 
However, when applying countermeasures, there are cases where the magnitude of the risk must also be considered. 
In general, the magnitude of risk is derived by multiplying the effect of risk and likelihood of risk \cite{ISO31000}. 
Since the risks identified in this study are those in which the ``harm'' of the risk actually materializes, their effect can be derived by each type of harm. 
On the other hand, the likelihood of risks cannot be derived easily, because likelihood of each risk is derived by risk factors, but the likelihood of each risk factor depends on the structure of system. 
So, derivation of magnitude is not easy. 
Deriving concrete methods of measures of magnitude is left as future work.

\subsection{Derivation of Risk Scenarios}
\label{SS-D-RiskScenario}
Based on this study, we were able to derive the sequence of events starting from the risk factors, leading to hazardous situations, and resulting in harm. 
From the perspective of countermeasures against these risks, deriving concrete risk scenarios can be desirable. 
A risk scenario is a more concrete representation of a sequence of risks, such as when a disinformation is intentionally entered into the system to cause a misjudgment (AMI), then the system generates an incorrect result (AMj) (The inputted disinformation is ``TRUE'' by the misjudgment), and the result is posted on social media, leading to the spread of inputted disinformation (AFr). 
This example corresponds to a specific scenario in which disinformation such as ``Yesterday's earthquake was artificially triggered'' is cleverly manipulated (by altering the input text) to be recognized as ``true.'' 
If the automated fact-checking system mistakenly returns a ``true'' result, 
the attacker uses this result as an endorsement to spread the disinformation, claiming, ``The theory that yesterday's earthquake was an artificial one is fact. Even fact-checking has confirmed it to be true.'' 
We believe the tables of risks in this paper serve as a useful reference when deriving such scenarios. 
We would like to explore methods as a future work that derive scenarios easier, for example, realize automation.

\subsection{Validity of the Case Study}
\label{SS-D-CaseStudy}
In this case study, we identified risks on the automated fact-checking system DEFAME. 
Even if risks were identified for a system other than DEFAME, the risks are expected to be similar to those identified in this case study. 
In other words, risks related to risk factors and hazardous situations can be identified in the system, external entities, and their boundaries, and risks related to harm can be identified at users and social media platforms. 
Because we think that the structures of risks on the other systems are the almost same as DEFAME, namely, risks arise at input data, internal AI malfunctions, or adversarial attacks, and the harm is suffered by users or third parties via social media. 
While this case study is qualitative and limited in scale, 
it demonstrates that the proposed taxonomy can be directly applied to real-world automated fact-checking system designs.

\subsection{Countermeasures Against Risks}
\label{SS-D-Countermeasure}
To address the risks associated with automated fact-checking identified in this paper, the following three approaches can be considered. 
Even though these countermeasures are used, 
all risks may not be eliminated on the system, however, those risks are expected to significantly reduce by using them. 
From the perspective of developers and distributors, understanding the potential risks in their systems and taking countermeasures against them are important for developing and distributing safe and secure systems.

\begin{enumerate}
\item Improving the accuracy of each entity's judgment

Improving the accuracy of the automated fact-checking system itself, and that of the external tools helps reduce the likelihood of risks occurring. 
Therefore, from a developer's perspective, improving accuracy serves as a risk mitigation. 
This measure is effective for both risk factors and hazardous situations lurking within the system.

\item Use of measures against adversarial attacks

Risks in Table \ref{T-GuideWord2} are caused by adversarial attacks. 
Therefore, using countermeasures against adversarial attacks helps reduce the probability of risks occurring. 
For example, implementing attack detection functions, improving robustness through adversarial training, and using countermeasures against poisoning are all effective risk mitigation approaches. 
These measures help protect against both entering adversarial inputs and attacks for entities in the system. 

\item Notification of risks to users and/or acceptance of risks under the contract

This paper has identified the risks of automated fact-checking systems. 
We think that presenting these risks to users and ensuring they understand risks before using the system can help mitigate these risks. 
For example, simply stating, ``Because this system uses AI, there is a possibility of misjudgments,'' may reduce the risk of users blindly accepting incorrect results and spreading them on social media. 
This countermeasure can prevent the spread of mis/disinformation. 
The identified risks in Table \ref{T-GuideWord1} and \ref{T-GuideWord2} can be used for determining what kind of notifications are effective for users.
Similarly, including a mention of these risks in user consent forms and contract terms is another potential countermeasure. 
This countermeasure is expected to encourage users to pay closer attention to data entry and the understanding of results, and to be more careful when posting results on social media. 
Additionally, if the system finds a user who is attacking the system, 
these consent and contract can be used as reason for banning that user. 

\end{enumerate}

\section{Conclusion}
\label{S-Conc}
In this paper, we categorized the risks on automated fact-checking systems. 
By utilizing existing risk assessment techniques and risk classifications, we categorized the risks considering the chronological sequence of risk occurring, namely, risk factors, hazardous situations, and harm. 
We identified four risk factors, four hazardous situations, and nine harm for normal use of the systems. 
For risks caused by adversarial attacks, we identified three risk factors, four hazardous situations, and eight harm. 
As a result, we found 32 types of risks totally associated with automated fact-checking systems. 
Using the identified risks as guide words, we conducted a case study to identify risks on DEFAME. 
As a result, we were able to identify risks on the DFD that could not be identified using STRIDE. 
We discussed the comprehensiveness of the risks, the magnitude of risks, derivation of risk scenarios, the validity of the case study, and countermeasures against the risks. 
The tables of risks identified in this paper are expected to be useful for developers and distributors of automated fact-checking systems when considering the risks of the systems they are developing and distributing. 
Since this case study was conducted virtually, to verify whether the identified risks actually manifest in real systems, 
and automated risk scenario derivation are marked as future works.

\section*{Acknowledgement}
Some of the findings in this paper are based on results obtained from a project, JPNP22007, commissioned by the New Energy and Industrial Technology Development Organization (NEDO).


\begin{thebibliography}{00}

\bibitem{Pizzagate}
BBC, ``The saga of 'Pizzagate': The fake story that shows how conspiracy theories spread,'' \url{https://www.bbc.com/news/blogs-trending-38156985}

\bibitem{earthquake-nhk}
Japan Broadcasting Corporation (NHK), ``Caution urged over Japan quake fake posts,''
\url{https://www3.nhk.or.jp/nhkworld/en/news/backstories/4758/}

\bibitem{politifact}
PolitiFact, \url{https://www.politifact.com/}

\bibitem{fullfact}
Full Fact, \url{https://fullfact.org/}

\bibitem{JFC}
Japan Fact-check Center, \url{https://www.factcheckcenter.jp/}

\bibitem{FJ-Press}
Fujitsu limited press release, `` Fujitsu to combat fake news in collaboration with leading Japanese organizations,'' \url{https://info.archives.global.fujitsu/global/about/resources/news/press-releases/2024/1016-01.html}

\bibitem{Pennycook}
Gordon Pennycook, David G. Rand, ``The Psychology of Fake News,'' Trends in Cognitive Sciences.

\bibitem{STRIDE}
Microsoft Corporation, ``Microsoft Threat Modeling Tool, '' \url{https://learn.microsoft.com/ja-jp/azure/security/develop/threat-modeling-tool-threats}

\bibitem{Nakov}
Preslav Nakov, David Corney, Maram Hasanain, Firoj Alam, Tamer Elsayed, Alberto Barr\'{o}n-Cede\~{n}o, Paolo Papotti, Shaden Shaar, Giovanni Da San Martino, ``Automated Fact-Checking for Assisting Human Fact-Checkers,''arXiv,2021, \url{https://arxiv.org/abs/2103.07769} 

\bibitem{Guo}
Zhijiang Guo, Michael Schlichtkrull, Andreas Vlachos, ''A Survey on Automated Fact-Checking,'' 
Transactions of the Association for Computational Linguistics, Volume 10. 

\bibitem{AISI-KA}
Japan AI Safety Institute, ``Known Attacks and Their Impacts on AI Systems,'' \url{https://aisi.go.jp/assets/pdf/Known_Attacks_and_Their_Impacts_on_AI_Systems_V2_EN.pdf}

\bibitem{poynter}
Baybars \"{O}rsek, ``Opinion | Structural problems with tech platforms prevent fact-checkers from focusing on harm and virality,'' 
\url{https://www.poynter.org/commentary/2024/structural-problems-with-tech-platfor} \url{ms-prevent-fact-checkers-from-focusin} \url{g-on-harm-and-virality/}

\bibitem{Tanaka}
H. Tanaka, M. Ide, J. Yajima, S. Onodera, K. Munakata, N. Yoshioka, ``Taxonomy of Generative AI Applications for Risk Assessment,'' 
2024 IEEE/ACM 3rd International Conference on AI Engineering - Software Engineering for AI (CAIN 2024), 2024. 

\bibitem{AISI-SG}
Japan AI Safety Institute, ``Guide to Evaluation Perspectives on AI Safety (version 1.10),'' \url{https://aisi.go.jp/assets/pdf/ai_safety_eval_v1.10_en.pdf}

\bibitem{kumamoto}
Japan Today, ``Man arrested for posting false tweet claiming lion on the loose after Kumamoto quake,'' 
\url{https://japantoday.com/category/crime/man-arrested-for-posting-false-tweet-} \url{claiming-lion-on-the-loose-after-kuma} \url{moto-quake}

\bibitem{DEFAME}
T. Graun, M. Rothermel, M. Rohrbach, A. Rohrbach, `` DEFAME: Dynamic Evidence-based FAct-checking with Multimodal Experts,'' arXiv, \url{https://arxiv.org/abs/2412.10510}

\bibitem{FacTool}
I-C. Chern, S. Chern, S. Chen, W. Yuan, K. Feng, C. Zhou, J. He, G. Neubig, P. Liu, 
``FacTool: Factuality Detection in Generative AI -- A Tool Augmented Framework for Multi-Task and Multi-Domain Scenarios,''
arXiV, 2023, \url{https://arxiv.org/abs/2307.13528}. 

\bibitem{ISO51}
International Organization for Standardization (ISO), ``ISO/IEC Guide 51: 2014, Safety aspects -- Guidelines for their inclusion in standards, '' \url{https://www.iso.org/standard/53940.html}

\bibitem{FTA}
Phillip J. Brooke, Richard F. Paige, ``Fault trees for security system design and analysis,'' Computer \& Security, Volume 22, Issue 3.

\bibitem{ISO31000}
International Organization for Standardization (ISO), ``IEC Guide 31010:2019, Risk management  -- Risk assessment techniques, '' \url{https://www.iso.org/standard/72140.html}


\end{thebibliography}
\end{document}